# Tainted Love: A Systematic Review of Online Romance Fraud

Alexander Bilz[0000-0002-0692-2482], Lynsay A. Shepherd[0000-0002-1082-1174], Graham I. Johnson[0000-0002-7610-862X]

School of Design and Informatics
Division of Cyber Security
Abertay University
Dundee, United Kingdom
{a.bilz, lynsay.shepherd, g.johnson}@abertay.ac.uk

## Abstract

Romance fraud involves cybercriminals engineering a romantic relationship on online dating platforms. It is a cruel form of cybercrime whereby victims are left heartbroken, often facing financial ruin. We characterise the literary landscape on romance fraud, advancing the understanding of researchers and practitioners by systematically reviewing and synthesising contemporary qualitative and quantitative evidence. The systematic review provides an overview of the field by establishing influencing factors of victimhood and exploring countermeasures for mitigating romance scams. We searched ten scholarly databases and websites using terms related to romance fraud. Studies identified were screened, and high-level metadata and findings were extracted, synthesised, and contrasted. The methodology followed the PRISMA guidelines: a total of 232 papers were screened. Eighty-two papers were assessed for eligibility, and 44 were included in the final analysis. Three main contributions were identified: profiles of romance scams, countermeasures for mitigating romance scams, and factors that predispose an individual to become a scammer or a victim. Despite a growing corpus of literature, the total number of empirical or experimental examinations remained limited. The paper concludes with avenues for future research and victimhood intervention strategies for practitioners, law enforcement, and industry.

**Keywords:** Online Dating, Romance Fraud, Romance Scams, Cybercrime, Human-Computer Interaction, Systematic Review.

## 1 Introduction and Background

Romance fraud is a form of social engineering which emerged in the early 2000s, whereby fraudsters create a fake profile on dating platforms and strike up a relationship with their potential victims, with the end goal of conning individuals out of money. The resulting damage to victims of romance fraud can be devastating; in addition to monetary loss, there is a huge emotional impact for users desperate for companionship. Romance fraud has become a growing problem, exacerbated by the COVID-19 pandemic, alongside other cybercrimes (Kemp et al., 2021; Lallie et al., 2021).

The execution of romance scams typically follows a multi-staged model (Whitty, 2013b). Generally, the scam involves the following steps: It starts with the scammer setting up an attractive profile on a dating site or social media network that either replicates a celebrity or figure of authority or portrays an entirely fictitious identity (Sorell and Whitty, 2019). Once the profile has been created, the scammer reaches out to potential victims and lures the prey into an online conversation (Rege, 2009). Soon after communication has been established, the conversation between the perpetrator and the target is moved to another channel, such as instant messaging, text messaging or email, outside the dating platform's control (Whitty, 2013a). This measure, a precaution to evade the detection mechanisms of dating sites, is commonly justified by appealing to intense emotions and exclusivity. The grooming phase follows, in which a strong, well-trusted, emotional relationship is established before the victim is asked for financial support to overcome a tragic or desperate situation (Whitty, 2013b). According to Whitty (2013a), monetary requests can take various forms, such as a small initial request to probe the victim's willingness to part with their money or an immediate large amount that's later negotiated down. Reasons for the monetary requests may include inheritance fees, plane tickets, visa fees, family emergencies or presents (Cross and Holt, 2021; Whitty, 2013a). The scam usually ends several months or years after the initial contact, either when the targets run out of money or realise they have been victimised (Cross, 2016a; Cross and Blackshaw, 2015).

A key difference between romance fraud and other financial dating scams lies in the scam's *modus operandi*. Romance scammers, as opposed to gold diggers, never seek to establish the relationship even temporarily, but merely use the perception of a relationship to take advantage of the victim (Thompson, 2016; Whitty, 2013b).

Victims of romance scams typically experience a "double hit" as they simultaneously lose significant financial means and are emotionally wounded due to the loss of the relationship (Whitty and Buchanan, 2012). Even though individual financial losses can be substantial, with recent reports as high as £300,000 in one case (BBC, 2021), emotional consequences have been estimated to be even more impactful than the perceived consequences of the financial losses (Modic and Anderson, 2015).

Despite the impact of romance fraud on society and the damage caused to individuals, no comprehensive systematic review has been conducted that coherently synthesises and then reflects on the literature in the field. Therefore, the objective was to answer three research questions that are pertinent to understanding romance scams:

- **RQ1:** What is the state-of-the-art approach for profiling, describing and characterising romance scams?
- **RQ2:** What underlying socio-demographic and psychological factors enable and foster the execution of a scam?
- **RQ3:** What countermeasures and mitigation techniques have been proposed to help identify romance fraud, thus limiting the risk of victimisation?

This research starts by analysing existing literature based on meta-level information, such as the type of literature (journal articles, conference proceedings or theses), the main contributions in line with three research questions, and authorship to establish an overview of the current literary landscape. Subsequently, the individual papers and their findings are presented, critically appraised, and contrasted to identify commonalities and discrepancies. This paper closes with an extensive discussion of gaps in the current knowledge and directions for future research.

## 2 Methodology

The following section presents the methodology used within this systematic review on romance fraud.

### 2.1 Protocol and Registration

We adopted the PRISMA (Preferred Reporting Items for Systematic Reviews and Meta-Analyses) technique and protocol to search, collect, and analyse relevant literature (Page et al., 2021). PRISMA provides a well-structured approach for reporting the work conducted and the findings of systematic reviews. This study aimed to provide a state-of-the-art view of the published literature on romance fraud and include contributions from different fields; thus, advanced statistical data processing was not feasible. Therefore, the following sections present our adaption of the PRISMA technique while closely adhering to the relevant checklist's items, such as the specification of the eligibility criteria, search strategy, and data collection process.

### 2.2 Eligibility Criteria

The primary inclusion criterion was that papers had to involve empirical research or experimental examinations that profile romance scams (including demographics), characterise the psychological aspects involved, or examine countermeasures and mitigations. Additionally, literature considered acceptable were peer-reviewed journal articles and book chapters, conference proceedings, theses, and dissertations. The choice to include grey literature (defined as electronic and print literature that institutions produce outside of the control of commercial publishers (Auger, 1998)) was deliberate, as it allowed to collect a more recent and comprehensive picture of the current research on romance fraud. The search was limited to English language papers available through the University's subscriptions.

Several articles were omitted as they featured unclear research methodologies or focused on other types of online dating fraud (e.g., identity fraud or e-whoring), which lack the double-hit effect unique to romance fraud. Also excluded were articles whose primary research was only weakly related to romance scams. In cases where the same findings were published twice (e.g., as a conference paper and journal article) as separately standing works, the less extensive paper was excluded.

## 2.3 Information Sources

Existing romance fraud research has seen contributions from various domains using varied research methods. To reflect the diverse and disparate nature of the contributions in this review, a wide range of information sources were searched for articles, including journal articles, conference proceedings, theses and dissertations and book chapters. The bibliographic search, conducted in October 2021, explored the following reputable and renowned publication databases: Scopus, PsycInfo, PsycArticles, Medline, Web of Science, ProQuest, ACM Digital Library, IEEE Digital Library, ScienceDirect and PubMed. Two additional sources were also utilised. Firstly, the reference lists of publications included in the full-text review were screened for relevant inclusions. Secondly, a Google Scholar alert was created using the search query detailed in section 2.4 to cover literature newly released within the month of the systematic literature review.

## 2.4 Search Strategy and Selection Process

The first step consisted of constructing a search string and querying the information sources for articles capable of answering the research questions outlined in Section 1. Due to the research questions' breadth and the methodology's exploratory nature, the search query was designed to capture the largest possible body of relevant literature. Therefore, the search query included an extensive range of euphemisms and synonyms commonly associated with romance fraud and refrained from using Boolean conditions to narrow down the number of articles at the time of the search. While it is acknowledged that this can lead to a larger number of irrelevant articles being returned, it was considered preferable over prematurely excluding articles from the overall scarce corpus of literature. The construction of the search query also meant it could be used across all databases without considering the unique technical limitations of the individual database searches, such as their capabilities of processing nested conditions and the number of supported search terms.

The final search query was as follows: *"romance fraud" OR "romance scam" OR "sweetheart swindle" OR "dating scam" OR "love scam" OR "relationship fraud" OR "relationship scam"*.

The selection process was designed with scientific scrutiny and reproducibility in mind. Therefore, all process steps were distinguished into separate tasks, and all intermediates and outcomes were documented. At first, the publication databases were searched using the final search query. Identified articles were imported into Covidence (2022), where the two-staged study selection was performed. In the first stage, the article's title and abstracts were screened according to the inclusion and exclusion criteria. Articles that were found to be eligible were subsequently screened once more based on their full text.

## 2.5   Data Collection Process and Data Extraction

A custom data extraction template was created within Covidence to ensure consistent and reliable data extraction from selected articles. The data items assessed by the data extraction form can be grouped into two distinct categories: high-level metadata relevant for characterising the field of the literature and in-depth findings that capture study-specific data pertinent to answering the research questions. All extracted data was validated a second time within Covidence to ensure the accuracy of the extraction. A complete list of all the variables captured by the data extraction process can be found in Table 1.

| | |
|---|---|
| **Metadata** | Information on the article: title, year, literature type (e.g., journal, conference paper, thesis), venue (name of the journal and conference), research field (e.g., computer science, linguistic, psychology, criminology), and Google Scholar citations |
| | Information on the authors: full name, affiliation (typical the university or research institute), location (country), and profession (e.g., academic, law enforcement, industry professional |
| | Funding information |
| | Contribution type: exploring/proposing profiles, methodologies, countermeasures, legal aspects, or discussion of open issues |
| | Study set-up: study objectives, research methodology (e.g., qualitative study, quantitative study, mixed study design, experimental), sample size, sample characteristics, country of research and approaches for gathering data (e.g., interviews, questionnaires, online forums, social media sites) |
| **Findings** | The perspective of the study (e.g., victims, scammers, website operators) |
| | Applied theories and frameworks |
| | Main Findings: approaches for profiling romance scams and their progression, demography and psychological characteristics of scammers and victims, and countermeasures and mitigations |
| | Contributions of the article |

*Table 1 Data items collected from the selected studies*

## 3   Results

The following section presents the main findings of the systematic review. At first, the study selection and study characteristics are introduced. Afterwards, the main findings are presented, grouped by their primary research contribution.

## 3.1 Study Selection

A total of 270 studies were identified across 10 databases and secondary sources (Table 2). Of these, 38 were removed as duplicates and 142 were excluded based on the title and abstract screening as they did not meet the inclusion and exclusion criteria (Figure 1). The remaining 90 articles were assessed in full text. Twelve were excluded because they not only referred to romance scams but failed to cover them coherently. Nine articles had research methodologies which were outwith the scope of our study (i.e., the papers utilised non-empirical research methods such as literature reviews and scoping reviews). An additional nine were removed as their full texts were not in English, and eight were not available in full text. Furthermore, another five articles were excluded because they had unclear research methodologies and three duplicated findings of a study by the same author, leaving 44 studies to be included in this review (Table 3).

| Search Source | Number of items before deduplication |
|---|---|
| Scopus | 69 |
| PsycInfo | 0 |
| Psycarticle | 0 |
| Medline | 0 |
| Web of Science | 44 |
| ProQuest | 39 |
| ACM Digital Library | 8 |
| IEEE Digital Library | 2 |
| ScienceDirect | 59 |
| PubMed | 12 |
| Manually Included | 37 |
| **Total number of articles** | **270** |

*Table 2 Search results by source*

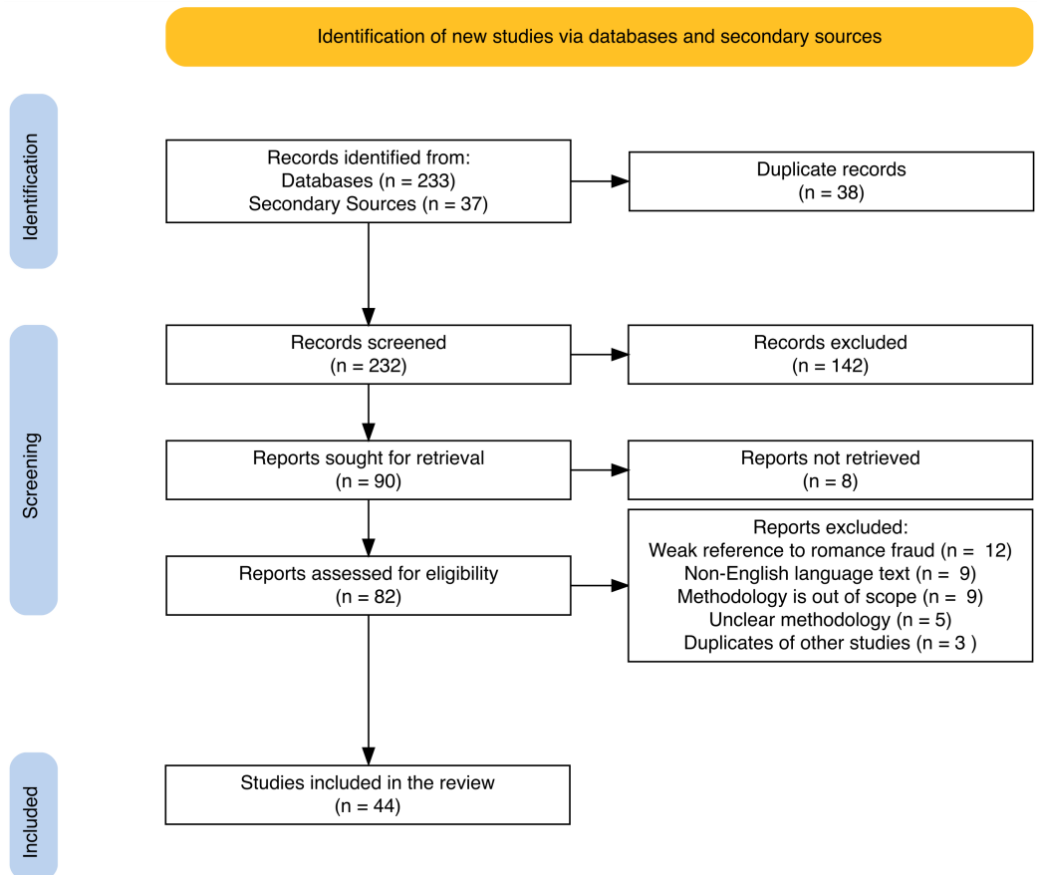

*Figure 1 PRISMA flow diagram for identifying articles related to romance fraud*

| Authors | Location | Research design | Methodology | Sample |
|---|---|---|---|---|
| Al-Rousan et al. (2020a) | USA Saudia Arabia | Quantitative | Machine learning | Not mentioned |
| Anesa (2020) | Italy | Qualitative | Linguistic Profiling | 26 online messages and 43 template messages |
| Barnor et al. (2020) | Ghana | Qualitative | Semi-structured interviews | 10 individuals engaged in romance fraud |
| Buchanan and Whitty (2014) | UK | Quantitative | Questionnaire | Study 1: 853 members of an online dating site Study 2: 397 members of a website to support romance scam victims |
| Buil-Gil and Zeng (2021) | UK | Quantitative | ARIMA modelling | Average of 4,166 respondents per month |

| Button et al. (2014) | UK | Qualitative | Interviews/Focus Groups | Study 1: 15 online fraud victims<br>Study 2: 6 focus groups with a further 48 online fraud victims<br>Study 3: 9 professional stakeholders. |
|---|---|---|---|---|
| Carter (2021) | UK | Qualitative | Discourse analysis | 1 conversation between a scammer and the victim |
| Cross et al. (2018) | Australia | Qualitative | Semi structured interviews | 21 victims of romance fraud |
| Cross (2019) | Australia | Qualitative | Semi-Structured interviews | First Round: 6 victims of romance fraud<br>Second Round: 7 victims of romance fraud |
| Cross and Holt (2021) | Australia USA | Mixed | Content analysis | 2478 complaints to Action Fraud |
| Cross and Layt (2021) | Australia | Qualitative | Content analysis | 509 victim reports to Scamwatch |
| De Jong (2019) | Netherlands | Quantitative | Machine learning | 4,154 profile images |
| Dickerson et al. (2020) | UK | Mixed | Pre-test and post-test experiment | 12 participants with who had used online dating platforms |
| Dreijers and Rudziša (2020) | Latvia | Qualitative | Linguistic Profiling | 7 letter sets |
| Edwards et al. (2018) | UK | Quantitative | Geocoding of IP addresses | 5,194 profiles of known scammers |
| Garrett (2014) | USA | Quantitative | Questionnaire | 110 Internet users |
| Graham (2021) | UK | Quantitative | Reverse image search | 240 ordinary profiles images, and 240 scammer profiles |
| He et al. (2021) | China Germany | Quantitative | Machine learning | 240 million posts, 320 million comments, and 33 million user profiles |

| Study | Country | Method | Technique | Sample |
|---|---|---|---|---|
| Huang et al. (2015) | UK | Mixed | Content analysis | 500,000 profiles of known scammers |
| Koon and Yoong (2017) | Malaysia | Qualitative | Linguistic Profiling | 21 email conversations |
| Kopp et al. (2015) | Australia | Qualitative | Content analysis | 37 profiles of known scammers |
| Kopp et al. (2016a) | Australia | Qualitative | Content analysis | 17 reports on an internet help forum |
| Kopp et al. (2016b) | Australia | Descriptive | Theory | No participants |
| Li et al. (2019) | Norway | Quantitative | Machine learning | 45 participants |
| Luu et al. (2017) | Australia | Quantitative | Questionnaire | 399 victims of romance fraud |
| Modic and Anderson (2015) | UK | Quantitative | Questionnaire | 6,609 individuals from the public |
| Obada-Obieh (2017) | Canada | Qualitative | Semi-Structured interviews | 10 general online daters |
| Offei et al. (2020) | Ghana USA | Quantitative | Questionnaire | 320 individuals engaged in romance fraud |
| Pan et al. (2010) | Australia | Mixed | Data mining | 5,481 profiles of known scammers |
| Pizzato et al. (2012) | Australia | Quantitative | Probabilistic modelling | 2,000,000 Expressions of interests from a dating site |
| Rege (2009) | USA | Qualitative | Document meta-analysis | 170 documents |
| Saad et al. (2018) | Malaysia | Quantitative | Study 1: Questionnaire Study 2: Association Learning | Study 1: 280 romance scam victims Study 2: 2,274 victim reporting to CCID |
| Shaari et al. (2019) | Malaysia | Qualitative | Content analysis | 60 online chats |
| Smeitink (2021) | Netherlands | Qualitative | Semi structured interview | 9 victims of romance fraud |

| Sorell and Whitty (2019) | Australia | Qualitative | Semi-structured interviews | 3 romance scam victims |
|---|---|---|---|---|
| Suarez-Tangil et al. (2019) | UK USA Australia | Quantitative | Machine learning | 14,720 ordinary profiles, and 5,402 scammer profiles |
| Webster and Drew (2017) | Australia | Qualitative | Semi-Structured interviews | 9 police officers |
| Whitty and Buchanan (2012) | UK | Quantitative | Questionnaire | 2028 individuals from the general public |
| Whitty (2013a) | UK | Qualitative | Study 1: Content analysis Study 2 + 3: Semi structured interview | Study 1: 200 posts from a public online support group; Study 2: 20 victims of romance fraud; Study 3: 1 SOCA officer |
| Whitty (2013b) | UK | Qualitative | Semi structured interview | 20 victims of romance fraud |
| Whitty and Buchanan (2016) | UK | Qualitative | Semi-structured interviews | 20 victims of romance scams or individuals who feel that they have been scammed |
| Whitty (2018) | Australia | Quantitative | Questionnaire | 11,780 Victims and non-victims |
| Whitty (2019) | UK | Quantitative | Questionnaire | 261 dating site and social media users |
| Whitty (2020) | Australia | Quantitative | Questionnaire | Study 1: 11,780 participants (10,723 non-victims, 1057 victims) Study 2: 531 participants (sub-set of study 1) |

*Table 3 Full list of articles included in the study*

## 3.2 Study Characteristics

This section provides an overview of the descriptive characteristics derived from the included articles. The final selection consisted of 30 journal articles, nine conference proceedings, four theses and dissertations, and a single book chapter. All studies were published between 2009 and 2021, with the majority of publications (n=25) being issued between 2018-2021 (Figure 2).

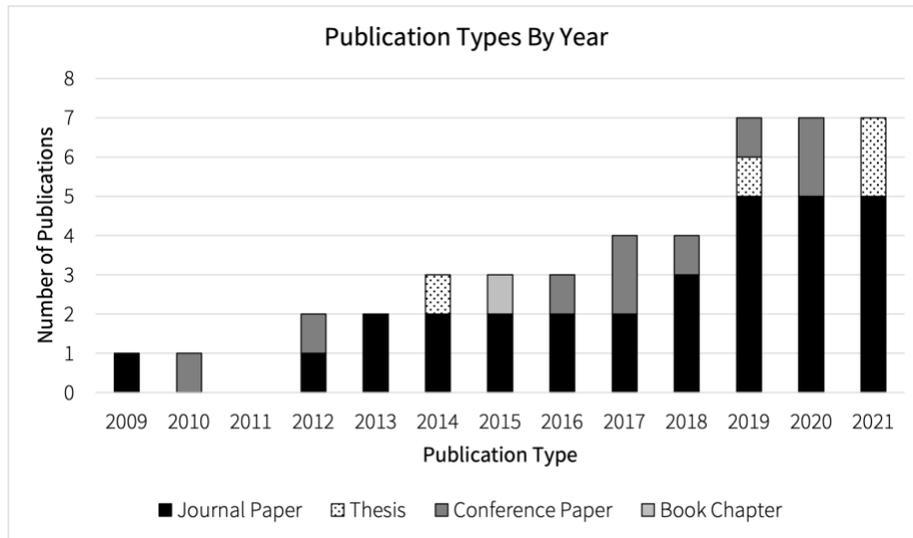

*Figure 2 Types of publications each year*

The contributors were distributed across thirteen different countries, based on the institutional affiliation listed (Figure 3).

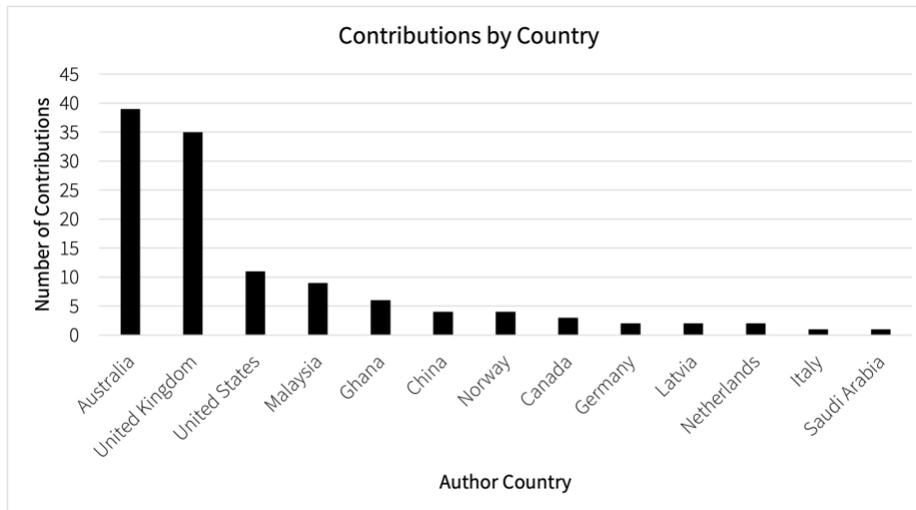

*Figure 3 Contributions by country*

In our analysis of the type of contribution present in the articles, we identified that the majority contributed to the profiling of romance scams and their manifestations (n=20), by the exploration of countermeasures for mitigating romance scams (n=14) and related factors of becoming a scammer or a victims (n=10).

The studies included in this review utilised a range of different research designs. Most prevalent were qualitative studies (n=21), primarily through semi-structured interviews, followed by quantitative research (n=19), frequently utilising online surveys.

## 3.3 Excluded Work

Articles were excluded if duplicate records found or there was no access to the full-text. Papers were also excluded if they were not written in English, or if they contained no relevance or a weak link to romance fraud. Furthermore, several papers were excluded if they did not feature a methodology of relevance to our review.

This incorporated papers with limited detail in the methodology in terms of participants and materials evaluated (Annadorai et al., 2018; Gregory and Nikiforova, 2012; Jimoh and Stephen, 2018). Several papers were also position-like pieces, featuring no empirical research (Gillespie, 2017, 2021; Steyerl, 2011; M. Whitty et al., 2017). Other papers featured a short review of existing literature, e.g., providing an overview of romance fraud, or linking to established theories (Cross, 2016b, 2020; Eseadi et al., 2021; Khader and Yun, 2017; Rege, 2013). Coluccia et al. (2020) conducted a scoping review of romance fraud's relational dynamics and psychological characteristics and identified twelve relevant studies. While such reviews contribute to the basic understanding of romance fraud, they do not deliver a detailed picture of the currently existing body of literature on the subject.

## 3.4 Profiling Romance Scams: Models, Taxonomies, Characteristics and Tactics

The following section presents the findings of studies that contributed to the profiling of romance scams through process models, taxonomies as well as discussions of characteristics and tactics prevalent in romance fraud.

### 3.4.1 Process Models of Romance Fraud

Explaining the progression of romance scams has been at the centre of academic interest. An early examination by Rege (2009) is one of the most referenced articles on this topic. Rege (2009) describes three distinct stages commonly found in romance scams. The scam begins with the creation of a fake profile and continues with the contact phase, which aims to establish trust and a strong bond between the scammer and the victim over a more extended period. The third and final stage entails one or multiple requests for money by describing a terrible or desperate situation.

The most frequently referenced model, however, was developed by Whitty (2013a), who identified five stages involved in the scam: being presented with an ideal profile, the grooming process (establishing the trust), the sting (request for money), sexual abuse and revelation. Subsequently, Whitty (2013b) proposed the Scammers Persuasive Model, a refined and extended version of her initial model, adding the motivation to find an ideal partner and a re-victimisation stage to the process flow. The work also acknowledges that not all victims experience all stages. In particular, the sexual abuse stage depends on the individual and the scam's setup and is no longer part of the immediate flow.

Smeitink (2021) developed a model based on romance fraud research in the Netherlands. The steps include: an individual being open to finding an ideal partner online, and the individual is presented with the ideal partner, the crisis, the continuation of the scam, confrontation, and the recovery phase as the chronological stages. Furthermore, Smeitink added the grooming phase as a dimension and the postmodern explanations of simulation, commodification and consumption as central to romance fraud. This change has been proposed to represent the nonlinear flow and highly dynamic nature of the grooming subprocess.

In contrast, Kopp et al. (2016b) argued that scammers engage a victim in two mutually supportive storylines: a relationship scam and a fund scam. While the former aimed at establishing a romantic relationship to develop commitment and trust, the latter sought to engineer the victim into transferring funds to the perpetrator.

In subsequent research, Kopp et al. (2016a) explained the transition between the storylines using the Transtheoretical Model's stages of change. They describe that the victim's interest is initially established by resonating with the personal love story. In the subsequent phase (action phase), gifts are exchanged, and communication is deepened by sharing personal similarities. Next follows the maintenance phase, a stage where victims experience strong romantic and happiness through regular communication. The relationship storyline ends with the termination of the relationship by the victim. Kopp et al. (2016a) explain that the fund story typically starts by introducing an uninvolved happening or a character. Subsequently, the story is expanded along with the relationship story until the money is requested. Money transfer becomes habitual during the maintenance phase, and transfers are seen as contributing to the relationship. The fund scam also ends with the victim terminating the relationship.

### 3.4.2 Characteristics of Romance Fraud

It is challenging to determine the actual number of occurrences of romance fraud due to the high likelihood of underreporting (Cross et al., 2018). Whitty and Buchanan (2012) estimated, based on survey data from 2,028 British adults, that 0.65% of all adults in the UK had already been scammed, and 2.28% know someone who had been victimised. The study by Modic and Anderson (2015) found similar numbers, as 1.6% of the survey participants indicated that they find lonely-hearts swindles plausible, and 1.5% indicated that they were likely to lose utility in the scam.

Whitty and Buchanan (2012) and Buchanan and Whitty (2014) indicate losses between £50 and £800,000 and median financial damages of £1,001 and £10,000 per victim. Similarly, Rege (2009) quotes losses of more than $3,000 for average victims.

Early qualitative research on romance fraud suggested that most scams originate from western parts of Africa, such as Nigeria, which were infamous for running fraudulent schemes (Rege, 2009; Whitty, 2013a). Later research conducted by Pan et al. (2010) confirmed these findings by highlighting that scammer profiles were frequently associated with IP address blocks assigned to the Lagos Region. Subsequent research by Edwards et al. (2018), which decoded IP addresses on a country level, identified that scams originated from Nigeria (30%), Ghana (13%), and Malaysia (11%), followed by South Africa (9%).

Furthermore, Rege (2009) established a typology of romance scammers, by describing the organisation and roles of the criminal networks operating romance scams. Rege points out that romance scammers' operations range from single individuals over close-tied small groups, to large-scale, loosely coupled networks. The six roles commonly found amongst scammers are: the organisers deciding on the nature and extent of the activities, the extenders in charge of expanding the network, and the executors who perform the fraudulent engagement with the victim. Additionally, there are also enforcers who compel victims to comply, money movers responsible for collecting and wiring funds and crossovers that are part of the scammer network and legitimate governmental, financial, or commercial entities.

### 3.4.3 Persuasive Tactics and Tricks

Researchers have explored various tactics that scammers use to lure victims into the scam and keep them engaged. Studies have performed linguistic profiling (Anesa, 2020; Dreijers and Rudziša, 2020; Koon and Yoong, 2017; Shaari et al., 2019), researched persuasion and repression tactics (Button et al., 2014; Carter, 2021; Cross et al., 2018; Kopp et al., 2015; Whitty, 2013b), and the usage of military identities to convey authority (Cross and Holt, 2021).

#### 3.4.3.1 Linguistic Profiling

Shaari et al. (2019) analysed written scam communications' structure and linguistic patterns, and identified three distinct stages: initial, pre-attraction, and hooked.

During the initial stage, the relationship is established. This phase is dominated by formal and polite communication, where scammers disclose details about their personality and successful career to gain the victim's trust. Similarly, Koon and Yoong (2017) confirm the prevalence of pseudo-self-disclosure to evoke trust, empathy, and sympathy by displaying the scammer as a successful, educated, affluent and elite individual. Anesa (2020) corroborates these findings and then adds that the self-presentation also attempts to evoke empathy and sympathy of the victim by sharing details about previous difficult experiences.

The second stage (pre-attraction), according to Shaari et al. (2019), aims at strengthening the online relationship. Common phrases during this stage include discussions about religion, expressions of attraction and phrases that prioritise the victim (Shaari et al., 2019). Again, Koon and Yoong (2017) confirm these findings and explain that scammers portray themselves as good, morally upright persons by using words and phrases with religious connotations. Koon and Young describe that the reasoning for the relationship is often stated as 'divine providence'. Other tricks used to underline a scammer's trust and credibility were expressions highlighting their online dating expertise, and their use of adjectives such as "honest" and "sincere".

Anesa (2020) notes that the appeal to strong emotions is used to establish a strong emotional bond. Extensive compliments regarding physical appearance and personality and messages, with flowery, cliché and hyperbolic elements that express the scammers' attraction and prioritisation of the victim, are also common (Anesa, 2020; Koon and Yoong, 2017). The expressions of strong attraction typically also contained requests to change to an alternative communication channel, a necessary step for evading the detection algorithms of online dating platforms (Anesa, 2020; Koon and Yoong, 2017).

The final phase (known as the hooked-phase) occurs from the mid-stage of the relationship until the money is requested. Shaari et al. (2019) and Koon and Yoong (2017) identified expressions that suggest feelings of indebtedness and gratitude for the received companionship. The monetary requests are constructed to evoke visceral responses from the victim while deflecting the deceptive nature of the message (Anesa, 2020). However, scammer writing styles could also be more aggressive, assertive, and occasionally offensive if victims fail to cooperate with monetary requests (Shaari et al., 2019).

Across all stages of the scam, Dreijers and Rudziša (2020) identified common micro-linguistic elements, including direct questions, flattering to keep the victim engaged, lyrical expressions and pet names to facilitate intimacy and bonding. Furthermore, elements appealing to core human drives to bond and feel were also used: buzzwords with romantic connotations, non-native English spelling and expressions, and sign-offs indicating a longing for the partner. On a macro level, Dreijers and Rudziša (2020) found four common stages: an initial situation, the reconnoitre (observation), delivery and the trickery stage. Similar to Shaari et al.'s (2019) model, the initial stage is about establishing communication and opening up, and the reconnoitre stage is concerned with collecting and requesting additional private information. In the third stage, the scammer fabricates an autobiography that is either full of integrity and success, or full of personal sorrow. Their model closes with obtaining personal details and/or requesting the money from the victim.

### 3.4.3.2   *Persuasion and Repression Tactics*

Button et al. (2014) researched reasons why victims fall for a range of online scams, including romance fraud. They discovered a variety of factors, such as fraud diversity, requests for modest sums of money from a large number of victims, authority and legitimacy, visceral appeals, embarrassing frauds, pressure and coercion, grooming, fraud conducted from afar, and different approaches. In particular, Whitty's (2013b) examination of romance fraud corroborates the usage of authority, highlighted through dating partners in authority, fast-moving relationships and visceral triggers, such as predicting future experiences and emotions experienced together, as contributing factors for motivational and cognitive errors. Additionally, Whitty identified appeals to urgency, reciprocation, norm activation, Cialdini's (2007) principles of commitment and consistency and love, liking and similarity, and addictive relationships.

Similarly, Carter (2021) establishes the practice of visceral responses and urgency through discourse analysis. Carter explains that at the start, the victim is provided with a false sense of power through visceral responses that indicate an emotional vulnerability of the scammer. Furthermore, Carter identified that scammers typically drop innocuous information about their financial and personal situation and revisit it throughout the communication to establish trust and believability. For the monetary request, Carter found that these were initially issued implicitly and later explicitly repeated, accompanied by stories of physical trauma and urgency to evoke the victim's visceral responses.

Kopp et al. (2015) introduces the concept of personal love stories as an integral factor for the success of the romance scam. The work indicates that scammers establish a relationship story that reflects the victim's perception of love and relationships, and tailor it throughout the scam to get the victim to believe that they have found their soulmate.

Carter (2021) adds the usage of isolation techniques, used to deflect the victim's suspicions about the fraudulent nature of the relationship, to discourage the victim from disclosing to family and friends. Scammers isolate the victim by implying loneliness and permanent loss of happiness if the victim attempts to end the relationship.

A study by Cross et al. (2018) considered the dynamics of romance fraud in comparison to concepts traditionally found in domestic violence research. The work highlights the creation of fear, isolation and monopolisation of the victim. Additionally, Cross states that scammers mistreat their victims through economic abuse, degradation, psychological destabilisation, Emotional or interpersonal withdrawal and contingent expressions of love. The economic abuse occurs through the request of money, while degradation manifests in verbal attacks. Psychological destabilisation is a technique, where victims are deliberately tricked into doubting their own decision-making skills and reality. Emotional or interpersonal withdrawal is a passive tactic that aims at increasing a victim's anxiety through temporally ceasing conversation, while contingent expressions of love build upon positive reinforcement to encourage the victim to comply.

### 3.4.3.3 Military Themes

In recent years, personas involving a military context have frequently been mentioned as distinct themes employed by scammers (Rege, 2009; Whitty, 2013b; Whitty and Buchanan, 2012). In military-themed scams, the perpetrators use profile photos of army or peacekeeping personnel and utilise a military context for creating the persona and storyline that projects authority (Anesa, 2020; Cross and Holt, 2021).

Based on the victims' reports from the Australian Consumer and Competition Commission, Cross and Holt (2021) identified that significant predictors for reporting a military theme were being female, younger in age, primary English speaker, and not living within Australia. Their work also identified that military scams typically start on social media platforms and are likely to involve the loss of personal details. However, it was found these types of scams showed typically did not show signs of abuse (e.g., threats or blackmail). These variables, however, were not found to be significant for indicating financial losses in the scam. Instead, males and victims with who already faced financial hardship were more at risk of monetary loss.

Cross and Holt (2021) explain that the usage of a military identity could either be incidental or central for requesting money from the victim. In cases where the army context was used to set up the profile, but not for the monetary demand, requests were excused as consignments/inheritances fees, health emergencies, criminal justice emergencies, assistance for family members, and a combination of these reasons. However, in cases where the whole military narrative was harnessed, money was requested for leave requests and resources, such as food and medicine. Cross and Holt stress that military-themed requests were commonly justified by the lack of access to personal financial means and the inability to communicate through voice or video for security and secrecy reasons.

### 3.4.4 Fraudulent Online Dating Profiles

Online dating profiles play a vital role in romance scams. It is typically the first point of contact that lures a victim into the scam and plays a crucial role in successfully executing the subsequent scam stages (Kopp et al., 2016a; Whitty, 2013a). Thus, the profiles and self-presentation of scammers have been subject to extensive research.

Pan et al. (2010) highlights that the majority (61%) of scammer profiles misrepresent their origin by claiming to be from the US. Follow-up studies confirm this e.g., Edwards (2018) notes that most scammers claimed to be from the US (63%), the UK (11%), or Germany (3%), as these are typically also the countries where their victims are located.

Differences between male and female scammer profiles have been characterised by average age and professions. Whitty (2013a) states that female profiles targeting heterosexual male victims in their later years typically portray themselves as no older than 30 years of age, and working in low-paying or non-professional jobs. For profiles targeting heterosexual females, the profiles were typically claiming to be as old as 50 but mostly younger than the victim. Occupations include professional jobs, entrepreneurial activities, or army ranks. For fake male homosexual profiles, Whitty (2013a) elaborates that these incorporate age ranges which are usually younger than 30 years of age and claim mixed professional status. Edwards et al. (2018) support these findings by indicating an average age of 30 for females and 50 for males. Similarly, Pan et al. (2010) identified that most females claimed to be between 20 and 29, and males stated ages between 40 and 49.

Disagreement exists on the prevalence of a particular gender linked to scammer profiles. While Pan et al. (2010) argues the predominance of young females (64.93%), Huang et al.'s (2015) analysis of flagged profiles from a Chinese dating site presents the majority (close to 80%) of romance swindlers as males. Similarly, Edwards et al. (2018) found that 64 % of all profiles were males.

Different findings also exist for the marital status indicated by the profiles. While Edwards et al. (2018) produce a convincing account for the existence (50%) of single status across all profiles, Huang et al. (2015) found a prevalence of widows (50%), followed by divorcees (33%), with single people only accounting for 15%. Whitty (2013a) indicates that the marital status is again dependent on the target characteristics, with male profiles typically posing as widowers with a child, whereas females commonly pose as singles.

Edwards et al. (2018) extended these perspectives and reason that the gender, ethnicity, marital status, and profession misrepresented in the profile can show unique characteristics based on the origin of the scam. Profiles originating from Nigeria, Malaysia, and South Africa primarily posed as males, while profiles from the Philippines, Ukraine, and Senegal presented themselves as females. Furthermore, some countries displayed occupations other than the predominant military and engineering themes. Purported Italian profiles often indicated real estate as an occupation, Philippines indicated sales, and Côte d'Ivoire identified primarily as a student. A similar tendency could be noted for countries that state ethnicities other than white. Profiles originating from Senegal and Côte d'Ivoire self-identified as black, while profiles from the Philippines stated mixed.

Kopp et al. (2015), performed a qualitative analysis of the profile descriptions of 37 fraudulent profiles, identified that the profile descriptions typically consist of a self-presentation, explanations of hobbies, motivations and the ideal partner. The study argues that male profiles describe themselves as masculine, wealthy, humorous, and God-fearing, while females portray confidence, financial independence, and occasionally sexually provocative suggestions. Men also frequently included personal tragedies in the profile, such as the loss of a loved one, suggesting their need for a caring, sympathising partner.

Additionally, two rather curious findings about scammer profiles were reported by Pan et al. (2010), who postulated that the fraudulent profiles were frequently associated with Yahoo's email service and descriptions that suggested a bisexual sexual orientation. One aspect both female and male profiles have in common is their use of attractive profile pictures commonly stolen from social media sites or modelling agencies (Cross and Layt, 2021; Rege, 2009; Whitty, 2013b).

### 3.4.5  Consequences of Romance Fraud

Romance fraud typically consists of a double hit, consisting of a financial and emotional impact, can be noted among victims (Rege, 2009; Whitty, 2013a). Buchanan and Whitty (2014) analysed the effects on the victim and showed that financial victims report higher emotional impact than non-financial victims, with 40% signifying they were 'very distressed over a long period'. Modic and Anderson (2015) add support to their findings. They determined lonely heart swindles as the fraud category with the highest emotional impact on the victim combined with reasonably high financial losses. Women were not just found to experience higher emotional impact, but also suffer higher financial losses than their male counterparts (Buchanan and Whitty, 2014).

The emotions that victims experience post-scam include shame, embarrassment, shock, anger, worry and stress, fear, and the feeling of being mentally raped (Whitty and Buchanan, 2016). In cases where sexual acts were performed on camera, victims also reported feelings akin to suffering sexual abuse. Cross (2019) states that victimisation also caters to other negative aspects of well-being, including physical and mental health problems, relationship problems, unemployment, homelessness, and suicidal thoughts. Other consequences are a diminished sense of self-worth and confidence, loss of trust in others and cutting social ties with former acquaintances (Whitty and Buchanan, 2016). Whitty and Buchanan (2016) further explain that for some victims, the perceived loss of their 'ideal partner' weighed heavier than the financial loss, as they perceived the fraudulent relationship as therapeutic and superior to any other relationship.

Whitty and Buchanan (2016) found that most victims struggle to cope with the experiences as they feel they cannot disclose to family, friends, or work colleagues out of fear of rejection and anger. This consequence leaves most victims in a stage of denial, where they are at particular risk of re-victimisation. Those victims who disclosed to peers frequently did not get the necessary support but experienced increased negative feelings of self-blame.

The notion of self-blame is common among romance scam victims, especially when victims cannot recognise the fraudulent nature of the relationship despite being warned by others (Sorell and Whitty, 2019). The work reasoned that the more victims disregard available counterevidence, such as warnings by third-party authorities, the more often they are victimised, and the more responsible they become for their losses. They emphasise that victims are generally prudent but fail to apply caution in their online relationships due to believing they are in love. Thus, they relax belief ethics and hyper-personal tendencies that make them vulnerable to pressure.

## 3.5 Influential Factors: Socio-Demographic, Experiential and Dispositional Characteristics of Scammers and Victims

Several articles (n=10) tended to focus on influential factors, such as scammers and victims' socio-demographic, experiential and dispositional characteristics which may increase the likelihood of them being involved in a romance fraud. These are discussed based on the typologies to which they contribute.

### 3.5.1 Typology of Victims

Garett (2014) indicates that active engagement finding a partner yields a higher likelihood of falling for a romance scam. Those who want a partner but do not actively engage in the search process showed higher monetary losses than other online daters. Another factor that significantly increases both the likelihood and severity is a focus on finding an international partner. Furthermore, the years of internet usage affect the probability and average losses. Garett (2014) demonstrates that respondents with one to five years of internet experience were particularly exposed.

Similarly, Saad et al. (2018) compared the influencing factors of victimisation among Malaysian romance scam victims. They found a moderate positive correlation between the level of cyber-fraud awareness and the risk of victimisation and a weak positive correlation for computer skills. These findings align with Garret's (2014) indications that a lack of scam awareness increases a victim's vulnerability and losses. Saad et al. (2018) identified age, education level and computer skills as additional influencing factors. Interestingly, using an Apriori algorithm, they found that well-educated, married women of Chinese and Malay ethnicity between the age of 25 and 45 have a high likelihood of victimisation. The authors did not elaborate on their finding that married people were more likely to experience victimhood than individuals who were not in a relationship.

A study by Whitty (2018), which compared socio-demographic and psychological characteristics of romance scam victims and non-victims in the UK, partly confirms these findings. Similar to Saad et al. (2018), Whitty suggests that well-educated women in the age group of 35–54 are at particular risk of victimisation, as this group is most likely to engage in online dating and have higher levels of disposable income. Additionally, psychological characteristics, including impulsivity, lack of self-control, addiction disposition, more trustworthiness and less kindness, were indicative of romance scam victimhood. In subsequent research, Whitty (2020) also analysed the psychological characteristics and socio-demographic differences across different cyber scams. Her study found that gender, education, impulsivity, locus of control, and neuroticism were significant predictors of cyber scam victimhood. Further analysis of personal characteristics by scam type identified that romance scam victims have higher impulsivity, education, and neuroticism scores compared to victims of other scams.

In assessing the effects of personality and psychological variables on the risk of being scammed and the emotional distress experienced after a scam, Buchanan and Whitty (2014) only identified idealisation, a sub-element of romantic beliefs, as the only significant factor but in practice, a limited predictor for victimhood. Idealisation is an individual's belief in the fulfilment of a perfect relationship. Other psychological characteristics, such as loneliness, personality and sensation, did not reach statistical significance.

### 3.5.2 Typology of Scammers

In contrast to the well-studied influencing factors for victims, research on perpetrators is still scarce. The following section presents two studies that primarily analysed the psychometric aspects of individuals conducting online romance fraud.

Barnor et al. (2020) rely on the Motivation Opportunity Ability (MOA) framework and the Rationalisation dimension of the Fraud Triangle Theory to explain male Ghanaian scammer's socio-economic drivers. They discovered that the main drivers of romance fraud are peer recruitment, poverty, unemployment, and low education and income. For the opportunity dimension, the conditions that allow or make it possible for people to engage in a behaviour, the study suggests that flaws in current law, its enforcement and lack of law enforcement capabilities are the primary enablers for committing romance fraud. The third and final dimension of the MOA framework, ability, can be described as necessary skills and proficiencies for completing a set task.

Based on their qualitative interviews, Barnor et al. (2020) ascertained the need for social abilities, such as teamwork and interactional skills, and technology-related capabilities to remain anonymous, like knowing how to use VPNs (Virtual Private Networks). However, no formal IT training or higher education was present among the interviewees. Similarly, Rege (2009) lists basic computing skills as required technical skills, alongside patience, social skills, and the ability to follow routines, as non-technical skills.

Most of the offenders interviewed by Barnor et al. (2020) rationalised their behaviour by stating that cybercrime is less severe than physical crimes like murder because their 'wealthy' victims are supposedly less affected by the loss of money. Others also justified their behaviour as retaliation for the harm caused to their forefathers during the colonisation times.

In a similar study, Offei et al. (2020) analysed Ghanaian individuals' justifications who execute online romance scams using Neutralisation and Denial of Risk Theory. They found that scammers only use the denial of victim aspect of the Neutralisation Theory to justify their deviant behaviour. Scammers alleviate the sole responsibility for the deception by reasoning that their extensive investments contribute to a romantic relationship with mutual responsibilities. Findings for the denial of risk theory were consistent with Barnor et al.'s (2020) findings on ability and rationalisation. Scammers are self-assured in their ability to avoid prosecution and rationalise their behaviour as less aggressive and dangerous than physical crimes. Rege's (2009) work aligns with these findings by explaining that scammers frequently resorted to denial of victim and denial of injury techniques and that anonymity is a crucial evasion technique for scammer organisations of all sizes.

## 3.6 Countermeasures and Mitigations

This systematic review identified two broad groups targeted by the research on countermeasures and mitigations: technology-centred and human-centred approaches. The technology-centred approaches can further be divided by whether a system seeks to detect fraudulent profiles actively or if the primary goal is to create a robust people recommender system. Similarly, the human-centred approaches can be grouped into two subfields. The first group entails research focusing on inherent safeguarding approaches, while the second group concentrates on externally provided intervention and coping strategies.

### 3.6.1 Research on Detection Mechanisms

The six papers focussing on detecting fraudulent online dating profiles and scammer behaviour using machine learning approaches can be grouped based on the input features used to distinguish between bogus and genuine profiles.

Al-Rousan et al. (2020b) proposed Social-Guard, an online application that checks if a given profile image portrays a celebrity and whether the photo could be located on another website. They emplyed Amazon's Rekognition API (Amazon, 2023) to identify celebrities and the

Google Vision API (Google, 2023) to reverse-search the images. In their study set-up, the authors validated their development by creating fake profiles on OkCupid (2023) and populating these with unseen and known profile photos of celebrities and non-celebrities. The authors claim that their tool achieves "satisfactory" effectiveness and established that the Google Vision API can be used to locate similar images.

De Jong (2019) utilised the reverse search capabilities of Google and Yandex to locate a labelled image on the internet and then extracted the text of the webpages, normalised it and created a feature set for the given images. The feature sets were subsequently used to train and evaluate multiple machine-learning classifiers. The author explained that his random forest classifier reached a maximum accuracy of 92.4 % at the expense of a high false-negative rate of 19.7 %. These results indicated that the proposed model is more likely to falsely classify a scammer's image as benign than classify a regular online dater as fraudulent.

Graham (2021) also employed reverse image search and text extraction of websites for identifying scammer profiles. Unlike De Jong (2019), however, the author did not train a machine-learning model. Instead, he developed a browser add-on communicating with a client-side Python application that reverse searches Google (2023), Yandex (2023) and TinEye (2023) and evaluates the identified web pages based on keywords in near real-time. This algorithm mimics a common safeguarding strategy of validating an online identity through reverse searches, outlined in section 3.6.3. By searching the retrieved websites for scam-related terms and the name displayed on the dating site, the add-on can identify fraudulent profiles and validate benign users' profiles based on other online presences. In an evaluation, the author tested the FaceCheck add-on on 40 profiles from OkCupid (2023) and Match.com (2023) and returned 35 inconclusive results, two name matches and three name mismatches. The name matches could be attributed to the profile owner upon manual validation. For the name mismatches, one profile could be identified as a false positive, one could not be validated, and the third was a potentially suspicious profile.

Suarez-Tangil et al. (2019) extended the number of factors and included static profile information, such as demographical data and profile descriptions in their evaluation. The authors utilised various machine learning classifiers based on the input variables and their completeness. For complete demographical data, such as age, occupation and gender, the team used a random forest classifier, while for incomplete profiles, a Naïve Bayes approach was used. The image classifier was implemented as an SVM (support vector machine) with a radial kernel and worked based on image captions detected based on the image. The authors used LibShortText's (Yu et. al, 2013) implementation of an SVM with a linear kernel to process the profile descriptors. The final prediction, of whether a profile was a scam or benign, was made by combining the votes of the three classifiers using an RBF (radial basis function) Ensemble Classifier. The authors elaborated that the best single classifier was the description-based SVM classifier followed by the demographic classifier. As for the demographic features, the occupation area stated in the profile was the most important feature. This finding is in line with observations presented in the discussion of profiling romance scams (see section 3.4.4). Overall, the authors proclaim an accuracy of 97 % for the ensemble classifier.

In a recent study, He et al. (2021) extended the static profile features utilised by Suarez-Tangil et al. (2019). They used dynamic features, such as user behaviours, to detect fraudulent profiles on a Chinese social media and dating app called Momo (陌陌) (2023). Their proposed content-based attention network, dubbed "DatingSec", consists of three layers: input, pattern extraction, and prediction. The input layer takes five types of features: profile features, community features, behavioural features, and topic distributions of posts and comments. The extraction layer utilises an MLP (Multi-Level Perceptron) for processing static information and multiple Bilateral-*long short-term memory* (Bi-LSTM) processing dynamic information. He et al. (2021) indicated that their combined approach reached a precision of 90.5 %, an F1 of 0.857 and an AUC (area under curve) score of 0.940. The optimal features identified in this study for detecting scammers were the static profile information (e.g., posts and comments), followed by behavioural features (e.g., posts per day, comments received, photos used in posts).

A different approach for validating an internet acquaintance was proposed by Li et al. (2019), who developed a Random Forest-based classifier that predicted the gender - based on stylometry and keystroke dynamics gathered through chat applications. The stylometry analysed four features: the average thinking time, the ratio of key deletions, the average number of letters in a word, and the average number of words within a message. Keystroke dynamics focused on the latency between keypresses and the duration of key presses. Using these six keystroke dynamics features, a person's gender could be predicted via a 15-minute chat conversation with an accuracy of 72 %. An evaluation based solely on the four stylometry features reached a prediction accuracy of 64 %. The authors explained that a combination of both approaches through a score-level fusion approach yielded a lower accuracy of 64 % and that feature-level fusions were not feasible. A significant limitation of this work is that detecting key dynamics, such as the latency between releasing the first and pressing the second key, required the installation of a custom-developed keylogger called BeLT (Mondal et al., 2017), as traditional chat clients do not record and transmit these. Therefore, these detection mechanisms must be integrated into applications where keystroke dynamics can be accurately recorded.

### 3.6.2 Research on Susceptibility of People Recommenders to Fraudulent Profiles

Pizzato et al. (2012) researched the sensitivity of recommender algorithms to scammer profiles. In an experiment, they assessed the recommendations of collaborative filtering, a hybrid (content-collaborative reciprocal plus content filter) and content-based "RECON" recommender algorithms. They identified that collaborative filtering and hybrid recommenders more often than regular recommend highly suspicious profiles, favouring popular and active profiles and candidates with high interaction and high reply rates.

### 3.6.3 Research on Intrinsically Motivated Safeguarding Strategies

Several pieces of literature (n=4) covering countermeasures and mitigation strategies focused on intrinsically motivated safeguarding approaches used by online daters, and their effectiveness at detecting online deceit.

Cross and Layts' (2021) large-scale survey showed that the most common action taken in response to suspicious profiles was to conduct an internet search of the online persona. Most frequent were reverse image searches or online searches based on addresses, phone numbers, personal details, and message content. In most cases, these created suspicion of communicating with a scammer due to warnings on scammer awareness sites, image reuse and identity mismatches. However, even in cases without search hits, participants were still suspicious due to the lack of a digital footprint, such as a social media profile.

In contrast, similar research by Obada-Obieh et al. (2017) indicated that only three out of ten participants searched for online social media presence as a precautionary method. Most frequently (six out of ten times), online daters looked for inconsistency in the dating profiles, overly attractive profiles, or complete duplicates to identify scammers, followed by reliance on their 'gut feelings'. Interviewees were particularly suspicious of profiles with profile pictures that were 'too cute' and descriptions that appeared to be hoaxes. This discrepancy in safeguarding strategies can potentially be explained based on the characteristics of the sample groups and the data sources. While Cross and Layt (2021) drew on reports from scam victims to Action Fraud (2023), Obada-Obieh et al. (2017) interviewed general online daters without the precondition of being involved in a scam.

Whitty (2019) examined the predictors of human accuracy in identifying fake and real profiles. The response time was identified as a critical predictor for accuracy, followed by the participants' previous experience with identifying fraudulent profiles. Longer response time and previous exposition to romance fraud significantly improved the ability to spot deception. In terms of personality and behavioural traits, the author found that strong romantic beliefs and low consideration of future consequences negatively affect predictors of accuracy, while high impulsivity improved accuracy. Whitty reasoned the positive effect of high impulsivity as the reliance on 'gut instincts', a safeguarding strategy also identified by Obada-Obieh et al. (2017). Overall, Whitty warned that distinguishing fake from genuine profiles remains a nontrivial task: on average, participants only classified 13.51 out of 40 profiles correctly.

While the former research covered the safeguarding techniques utilised by individuals, Luu et al. (2017) examined the factors and processes influencing the adoption of protection mechanisms, such as checks of identities or reporting of users, using the Protection Motivation Theory (Rogers, 1983). They identified that adopting protection mechanisms was mainly influenced by an individual's coping appraisal, response efficacy, and self-efficacy factors. In contrast, response cost played a minor role. The authors discovered that perceived vulnerability and perceived severity were most influential for the threat appraisal. In particular, thorough knowledge of the harm caused by romance fraud motivated individuals to use protection mechanisms.

### 3.6.4 Research on Externally Stimulated Intervention and Coping Strategies

The externally stimulated intervention and coping strategies can be distinguished based on the scam phase in which they intervene, either from the grooming phase onwards or after the scam. Dickerson et al. (2020) proposed targeted and dynamic warning messages displayed to potential victims when a trigger situation, such as a request for money by a potential scammer, occurs. Their study identified that contextualised warning messages helped increase user awareness and decrease risky behaviour. Furthermore, warning messages contextualised advice on how to spot scammers made available by dating sites and public entities. While none of the participants in the control group interacted with the educational information, all the participants in the experimental group engaged with the training material referenced by the warning messages.

Proactive policing strategies were another target of research. Webster and Drew (2017) explored the experiences of law enforcement officers involved in an early intervention model by the Queensland Police Service. They identified that proactive policing yielded the first promising results through qualitative interviews, particularly when victims were already questioning their relationship. However, the collected data also highlighted challenges, such as low success rates, lack of appropriate training to cope with emotional victims, and the need for a victim selection process that considers the correct timing of interventions.

Research by Cross (2019) examined the motivations, expectations, and actual experiences of individuals who joined police-run peer support groups after being targeted by scammers. Significant reasons for joining a peer support group were the interactions with like-minded, a sense of community, the expected release of the burden suffered by the victim, the prospect of giving and receiving help, and the sheer absence of other means of support, such as family and friends. Unlike the motivations, which were largely optimistic, the actual experiences were mixed. Most participants valued sharing their own story with others and subsequent gain of solidarity and acknowledgement, the ability to learn how they have been victimised or build a friendship. However, others acknowledged diverse challenges, such as lack of engagement in the group, logistical problems, high participant fluctuation rates of the group members, expectation mismatches, and perceived tension between financial and non-financial victims.

# 4 Discussion

This section discusses the main findings and limitations of existing studies, and future research directions will be presented.

## 4.1 Study Characteristics

After applying a set of selection criteria, 44 studies were included in this review. The majority (n=20) primarily focused on profiling romance scams and their manifestations, followed by 14 exploring countermeasures for mitigating them, and ten evaluating influencing factors of becoming scammers or victims. Contributions employed a wide range of research methodologies and frameworks, which made direct comparisons of individual findings and exploration of the field more challenging due to the limited homogeneity between these studies. Furthermore, many studies focused primarily on online crime research rather than romance fraud specifically, requiring a degree of caution when comparing and contrasting findings.

Evaluation of the research methodologies showed that qualitative and quantitative methods were almost similarly prevalent. Most qualitative studies contributed to the profiling of romance scams, while quantitative studies primarily shaped the understanding of countermeasures and influencing factors. Unsurprisingly, the sample size between both methods greatly varied.

Across all studies, the complexity of acquiring novel, accurate and complete primary data on the subject were acknowledged. Multiple authors, from different fields, voiced the challenge of collecting reliable data due to the underreporting of the crime, ethical and legal challenges, and the sensitive nature of the crime (e.g., de Jong, 2019; Rege, 2009). In particular, the collection of long-ranging dynamic information, such as complete chat logs covering all stages of the scam, starting from the message of interest on the dating site to the money transfer negotiated off-platform, was acknowledged as challenging (Carter, 2021). This challenge might also explain the frequent usage of questionnaires and surveys for collecting relevant data.
Additionally, multiple studies were based on the same data, while others did not fully disclose their data sources. Pairs of studies that were based on the same initial data collection are Cross and Holt (2021) and Cross and Layt (2021), as well as Whitty (2018) and Whitty (2020).

Even when datasets differed, commonalities could be identified based on the origin of the data. Multiple studies relied on data from datingnmore.com (2021), a self-proclaimed scammer-free online dating site, when analysing known genuine profiles. Others relied on data from scamdigger.com (2021), when examining scammers profiles (de Jong, 2019; Edwards et al., 2018; Graham, 2021; Pan et al., 2010; Suarez-Tangil et al., 2019). As both sites are publicly accessible and allow for easy access to a large pool of data, it is no surprise that three out of five studies that incorporated machine learning utilised at least one of these sites (de Jong, 2019; Graham, 2021; Suarez-Tangil et al., 2019). However, the firm reliance on these two data sources might lead to incorrect assumptions, as datingnmore.com primarily caters to an older audience and, therefore, might not represent the general dating population.

One way of furthering work on spotting scammers is through an extended collaboration between academics, online-dating providers, and law enforcement to identify legal and ethical ways to access the relevant data. Existing studies that utilise corporations with dating sites, such as those conducted by Huang et al. (2015) on an unnamed Chinese dating site and by He et al. (2021) on the Momo (2023) app, indicate promising findings.

## 4.2 Profiling Romance Scams

Considerable research contributions have been made in characterising the prevalence and impact of romance scams. In terms of experienced financial losses, studies have found losses between £50 and £800,000 and medians of £1,001 to £10,000 per victim (Buchanan and Whitty, 2014). As these numbers are frequently based on reports to anti-fraud agencies such as Action Fraud, the actual numbers are likely to be higher due to many unrecorded cases.

Origins of the scam have also been explored in several papers. Studies by Pan et al. (2010) and Edwards et al. (2018) confirmed that most romance fraud originates from the West African countries of Nigeria and Ghana. However, Edwards et al. also showed that a considerable number of scams also originate in Malaysia (11%), Turkey (3%), the Philippines (2%) and Russia (1.5%). Research on the scams originating from these parts of the world is still scarce, despite initial indications of differences in the scam's execution, and the groups they target (Edwards et al., 2018; Garrett, 2014). Thus, more research is needed that addresses romance scams by region to identify unique characteristics, commonalities and differences in the execution of the fraud.

Studies concerned with the persuasive tactics used by scammers achieve a general consensus among those in this review. Scammers utilise a range of persuasive tactics and linguistic devices to evoke the desired emotion in a particular scam stage. Tactics include visceral appeals, the creation of urgency, fast-moving relationships, appeals to strong emotions, and even isolation and monopolisation. Anesa (2020) argued that most tactics are particularly effective as these scam victims into processing clues peripherally, according to the Elaboration Likelihood Model by Petty and Cacioppo (1984). An improved understanding of how individuals process messages from scammers could provide an insight as to why individuals fall for such deceit and how it can be prevented.

The literature on the creation of fraudulent profiles has once again identified both similarities and discrepancies. While the frequent use of high-quality, stolen photos has been supported by the literature, the prevalence of specific scammer profiles is under debate. Recent findings suggest that significant differences in the proclaimed demographics and personality traits exist based on the target population and on the origin of the profile (Edwards et al., 2018; Whitty, 2013a). On average, male profiles stated higher age ranges and more prestigious professions than their female counterparts.

Finally, research on the consequences experienced confirmed the preconceived idea of a double hit that affects victims both financially and emotionally. Victims have reported feelings, including shame, embarrassment, shock, anger, worry, and physical- and mental health problems (Cross, 2019; Whitty and Buchanan, 2016). Other casualties also stated losing an "ideal", almost therapeutic, relationship (Whitty and Buchanan, 2016). Common themes were the struggle to cope with their experiences, due to a lack of support from peers, and a strong notion of self-blame.

### 4.3   Influencing Factors

Despite frequent contributions (n=8) to establishing influencing factors that increase the likelihood of victimhood, few factors are statistically significant (Garrett, 2014; Whitty, 2018). The indicative ones, such as gender, education, age, active engagement in online dating, desire to find an international partner, impulsivity, locus of control, and neuroticism, are not free from contradiction. In particular, gender has been highly controversial. While the studies by Whitty (2018) and Shaari et al. (2019) suggest that women are more at risk, Rege (2009) reasons that a romance scam is gender agnostic. Although, Whitty (2020) later suggests that men are more prone to fall for romance scams than women.

A better understanding of the influencing factors is needed, as it could help to tailor law enforcement efforts and improve intervention success (Webster and Drew, 2017). Hence, we suggest that future research disaggregates the available data by influencing factors to identify vulnerable groups to a particular scam type. Factors that should be considered are pre-dispositional factors, such as previous experiences in relationships, sexual orientation, physical- and mental health problems, and the financial situation prior to the scam. Additionally, such high-level data means it can be difficult to establish links between victim characteristics and the type of scammer profile they find engaging.

Saad et al. (2018) conducted a study based on association rule learning which identified four specific profile categories of victims in Malaysia. Despite these novel findings, the work has multiple shortcomings. Firstly, it is unclear why the type of fraud was defined as parcel scams in the associative learning rules, even though the document was concerned with romance scams. Secondly, the authors did not discuss their finding that married people are at particular risk of victimisation, which is seemingly counterintuitive.

Regarding profiling fraudsters, commonalities in the motivation and neutralisation techniques were identified, as scammers are primarily driven by the prospect of financial gain and neutralise their responsibility through the existence of the victim (Barnor et al., 2020; Offei et al., 2020). However, beyond that, very little is known about the demographics and psychological insights of the offenders.

## 4.4 Countermeasures and Mitigations

Regarding technical countermeasures, six studies were identified. The features considered were - profile images (Al-Rousan et al., 2020b; de Jong, 2019; Graham, 2021; Suarez-Tangil et al., 2019), the profile descriptions and demographics (He et al., 2021; Suarez-Tangil et al., 2019), keystroke patterns (Li et al., 2019), as well as the behaviour showcased by a scammer (He et al., 2021). Different data sets utilised in the study may influence the performance of the proposed countermeasure. A possible solution is the publication of a standardised dataset with known scammers and benign profiles. The publication of structured, machine-legible data could also enable a streamlined development of technical countermeasures, as current research (Graham, 2021; Suarez-Tangil et al., 2019) scraped unstructured and semi-structured data from websites like datingnmore.com and scamdigger.com. The publication of structured data on fraudulent profiles through an API, similar to existing solutions for managing phishing submissions, would also support the rapid integration into end-user-facing solutions, like browsers and chat clients.

It was noted that ensemble classifiers outperformed singular classifiers. Works by Suarez-Tangil et al. (2019) and He et al. (2021) independently confirmed the benefit of using an ensemble classifier that combines predictions of the single classifiers that make up the fraud detection system.

One limitation that all studies had in common is that they, the countermeasures, are rarely viable for end-users to utilise during online dating. The closest work to an integrated solution, a browser add-on proposed by Graham (2021), still required a locally running Python server to perform the reverse image search. Different applications that must run simultaneously might severely inhibit the adoption among online daters, and would be most unsuitable for non-technical users.

The analysis surrounding intrinsically motivated safeguarding strategies by Luu et al. (2017) showed that an individual's coping appraisal and the factors of response efficacy and self-efficacy significantly influence the adoption of safeguarding techniques when online dating (even more than threat appraisal). Similarly, Whitty (2019) reported that previous experience in online dating deception improves the ability to spot deceit. Therefore, training and awareness resources must be well disseminated on online dating sites and strengthen the individuals' belief that they can protect themselves from financial losses and emotional harm. Current guidance around online dating often focuses on listing risky behaviours, rather than persuading individuals to consider negative outcomes associated with romance fraud, and the individual's ability to cope. Therefore, we endorse Whitty's (2019) call to develop interactive awareness and training material that teaches potential victims how to detect and prevent romance scams. Similar methods utilised to educate employees about phishing have indicated improved confidence levels among participants (Cj et al., 2018).

As romance fraud is still a heavily stigmatised crime, few victims have been able to disclose their experiences and receive support (Cross, 2019; Whitty and Buchanan, 2016). While the research concerned with externally stimulated intervention approaches, such as peer support groups and active policing, has displayed some success, more work is needed to support victims (Cross, 2019; Webster and Drew, 2017). In particular, law enforcement plays a key role, as they are typically the first point of contact after a victim realised that they have been defrauded or even while victims are still caught in the scam in the case of active policing (Webster and Drew, 2017; Whitty and Buchanan, 2016). Hence, law enforcement personnel working with romance fraud victims need to be well trained in handling the traumatised, potentially disbelieving, and vulnerable victims, and inform victims about professional counselling and support.

## 4.5 Further Considerations

When looking at the current proposed technical countermeasures, it is clear that these primarily target the dating platform. However, based on the profiling of romance scams and the analysis of persuasive tactics, we have explained that scammers swiftly move the communication away from the first contact point to minimise their risk of getting blocked. This action, however, reduces the amount of behavioural data, such as exchange chat messages, that can be collected from the platform where the initial contact occurred. Therefore, we recommend that future research be conducted agnostically to changes in communication platforms.

Dating and relationships are highly delicate aspects of people's lives. Thus, we see a strong need for ethical and legal considerations when conducting research that might infringe upon the privacy of unsuspicious online daters. For future work in this space, it will be necessary to establish stringent policies that govern the secure collection, handling, storage, and deletion of data.

This review was able to identify contributions from a range of backgrounds, disciplines, utilising various frameworks and approaches. In order to harness the full power of the different research fields, such as psychology, linguistics and computer science, more interdisciplinary work is required, as has already been called for by Sorell and Whitty (2019). In particular, the development of advanced technical countermeasures, such as machine learning models, needs to be well grounded in established theory. Factors that come into play are the characteristics and processes of romance fraud, persuasive tactics, linguistic patterns, and influencing and propensity factors of becoming a victim. As this state of the art of review on romance fraud has demonstrated, this field offers ample opportunity for nurturing interdisciplinary research.

Ultimately, improved countermeasures and mitigation strategies can help prevent romance fraud. However, as reasoned by Barnor et al. (2020) and Offei et al. (2020), more effective law enforcement is also needed to apprehend perpetrators in countries where scams originate.

# 5 Conclusion

Romance fraud constitutes an ever-increasing challenge for online daters, dating platform providers, and law enforcement. In particular, the COVID-19 pandemic with its lockdown restrictions may have contributed to increases in victimisation. This systematic review has examined the major findings of studies in the field. It assessed the approaches for profiling romance fraud, the socio-demographic and psychological factors of persons involved in the scam, and countermeasures and mitigation techniques.

Our comprehensive analysis of the salient study characteristics showed that the field has received considerable interest, especially during the last three years, but the total number of relevant studies remained small (n=44). Most studies in our review contributed to the profiling of romance fraud by establishing process models to describe the progression of the scam, or analysing linguistic patterns and persuasion techniques employed by the scammers when interacting with their victims. Additionally, studies examined scammers' online dating profiles for commonalities, such as occupations, marital status, and age. These findings can aid in developing improved detection and mitigation techniques and seed the development of practical training and awareness programmes.

Countermeasures and mitigation techniques have also received much attention. Various studies proposed that machine learning classifiers can effectively detect fraudulent profiles on online dating sites by considering features, such as the profile characteristics, the profile image and behaviour. However, as indicated in section 4.4, the absence of common datasets for training and testing and reporting standards severely limits development and assessment.

Less commonly found within the review were contributions on the socio-demographic and psychological factors, including age, gender, and education of victims and scammers. Whilst diverse characteristics attributing to victimisation have been considered, few were statistically significant. As romance fraud is still a relatively new and multifaceted research field, it is not surprising that there is limited attention to the factors that increase the risk of victimisation.

Whilst valuable contributions have already been made in the field, an additional cross-organisational collaboration between researchers, platform providers and law enforcement will be required to address current issues.

Romance fraud will remain an arms race between cybercriminals and bona fide online daters. As new attacks emerge and existing ones evolve, future research will have to keep pace with the perpetrators and develop new ways to prevent attacks and to provide engaging training and awareness programs and effective incident detection and response.

## Funding

This research was funded by the Scottish Funding Council.

## Conflict of Interest

The authors report there are no competing interests to declare. The funders had no involvement in the study's design, data collection, analysis, interpretation, manuscript writing, or publication of the findings.

## Acknowledgements

The authors would like to thank the Scottish Funding Council and Abertay University Graduate School.